\documentclass[twocolumn,showpacs,preprintnumbers,amsmath,amssymb,groupedaddress]{revtex4}

\usepackage{graphicx}

\begin{document}

\bibliographystyle{prsty}
\title{Orbital ordering in La$_{0.5}$Sr$_{1.5}$MnO$_4$ studied by model Hartree-Fock calculation}

\author{K. Ebata}
\affiliation{Department of Complexity Science and Engineering and Department of Physics, University of Tokyo, 
5-1-5 Kashiwanoha, Kashiwashi, Chiba, 277-8561, Japan}
\author{T. Mizokawa}
\affiliation{Department of Complexity Science and Engineering and Department of Physics, University of Tokyo, 
5-1-5 Kashiwanoha, Kashiwashi, Chiba, 277-8561, Japan}
\author{A. Fujimori}
\affiliation{Department of Complexity Science and Engineering and Department of Physics, University of Tokyo, 
5-1-5 Kashiwanoha, Kashiwashi, Chiba, 277-8561, Japan}
\date{\today}

\begin{abstract}
We have investigated orbital ordering in the half-doped manganite La$_{0.5}$Sr$_{1.5}$MnO$_4$, which displays spin, charge and orbital ordering, by means of unrestricted Hartree-Fock calculations on the multiband $p$-$d$ model. From recent experiment, it has become clear that La$_{0.5}$Sr$_{1.5}$MnO$_4$ exhibits a cross-type $(z^2-x^2/y^2-z^2)$ orbital ordering rather than the widely believed rod-type $(3x^2-r^2/3y^2-r^2)$ orbital ordering. The calculation reveals that cross-type $(z^2-x^2/y^2-z^2)$ orbital ordering results from an effect of in-plane distortion as well as from the relatively long out-of-plane Mn-O distance. For the ``Mn$^{4+}$" site, it is shown that the elongation along the $c$-axis of the MnO$_6$ octahedra leads to an anisotropic charge distribution rather than the isotropic one.
\end{abstract}


\maketitle

Perovskite-type transition-metal oxides have attracted much attention due to their wide variety of magnetic, electrical and structural properties. In particular, manganites are extensively studied because of their rich physical properties such as colossal magnetoresistance (CMR) and spin, charge and orbital ordering \cite{Tokura}. Historically, their electronic states have been understood in terms of a double-exchange (DE) mechanism with localized $t_{2g}$ electrons and itinerant $e_g$ electrons \cite{Zener, Anderson}. Also, it has been recognized that the $e_g$ orbitals are strongly hybridized with the oxygen $p$ orbitals, 
and participate in the cooperative Jahn-Teller (JT) distortion of the MnO$_6$ octahedra \cite{Millis1, Millis2}. The degeneracy of the $e_g$ level is lifted by the JT distortion and the resulting orbital polarization at the Mn$^{3+}$ sites can be accompanied by certain types of antiferromagnetic (AFM) spin ordering. Half-doped manganites such as La$_{0.5}$Ca$_{0.5}$MnO$_3$ and La$_{0.5}$Sr$_{1.5}$MnO$_4$ 
exhibit a so-called CE-type AFM ordering, as shown in Fig. 1. The magnetic moments of Mn on a zigzag chain are coupled ferromagnetically, whereas the neighboring chains are coupled antiferromagnetically. The charge and orbital ordering in the single-layered perovskite La$_{0.5}$Sr$_{1.5}$MnO$_4$ and three-dimensional perovskite Nd$_{0.5}$Sr$_{0.5}$MnO$_3$ were investigated using resonant x-ray diffraction and an alternating Mn$^{3+}$/Mn$^{4+}$ pattern were observed directly \cite{Murakami, Nakamura}. Mizokawa and Fujimori \cite{Mizokawa1} showed that JT distortion consistent with orbital ordering plays an important role in stabilizing the CE-type AFM state by means of unrestricted Hartree-Fock calculations. The orbital ordering of the $e_g$ electrons had been thought of as a rod-type $(3x^2-r^2/3y^2-r^2)$ orbital ordering, as shown in Fig. 1(a). More recently, however, from the studies of liner dichroism in the Mn 2$p$-edge x-ray absorption \cite{Huang} and resonant soft x-ray scattering at the Mn 2$p$ edge \cite{wilkins} of the La$_{0.5}$Sr$_{1.5}$MnO$_4$, where $T_{CO} \simeq$ 217 K and $T_N \simeq$ 110 K \cite{Sternlieb}, it was demonstrated that $e_g$ electrons exhibit predominantly cross-type $(z^2-x^2/y^2-z^2)$ orbital ordering as shown in Fig. 1(b).
\begin{figure}
\begin{center}
\includegraphics[width=4cm]{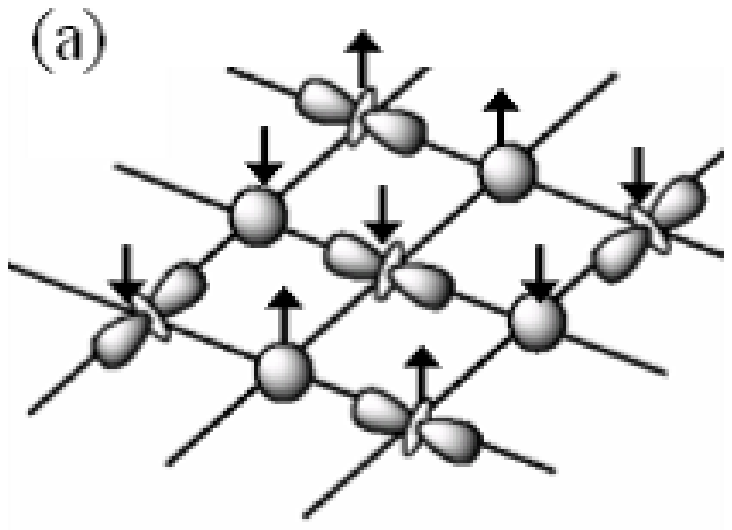}
\includegraphics[width=4cm]{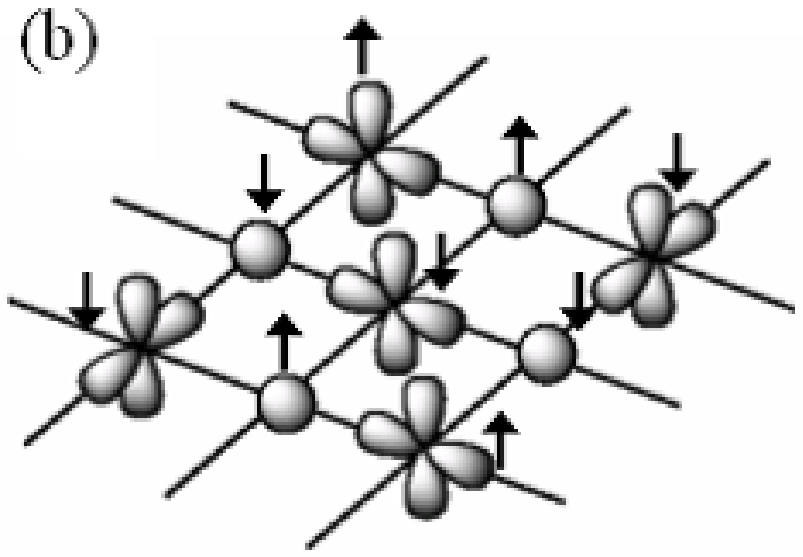}
\includegraphics[width=3cm]{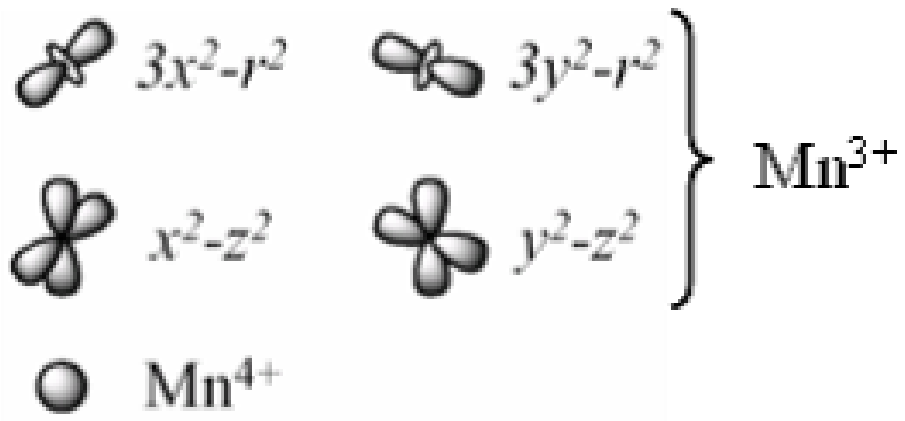}
\caption{Models of orbital ordering in the MnO$_2$ plane. (a) Rod-type $(3x^2-r^2/3y^2-r^2)$ orbital ordering; (b) Cross-type $(z^2-x^2/y^2-z^2)$ orbital ordering \cite{Huang}. Also shown is the CE-type AFM ordering by arrows.}
\label{fig1}
\end{center}
\end{figure}
\begin{figure}
\begin{center}
\includegraphics[width=7cm]{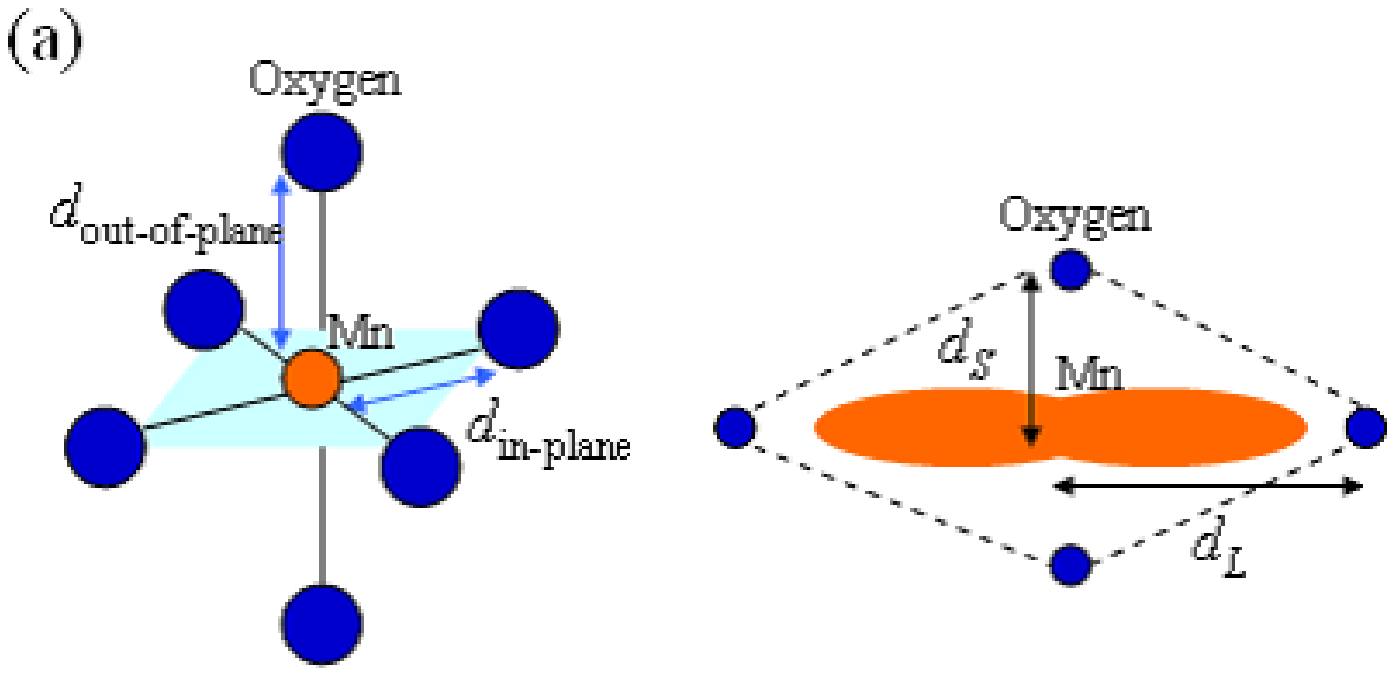}
\includegraphics[width=6cm]{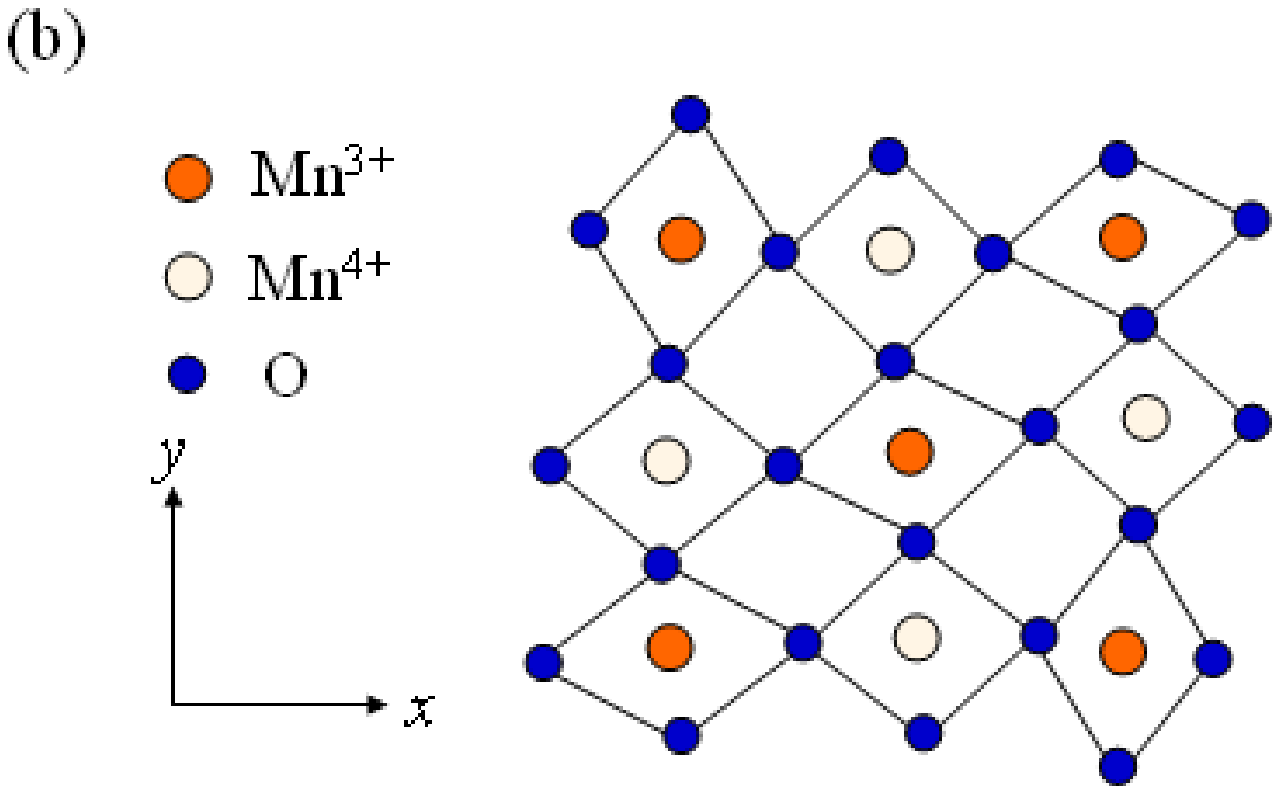}
\caption{(Color online) (a) Two parameters representing of lattice distortion at the Mn$^{3+}$, that is, the out-of-plane elongation of the MnO$_6$ octahedra represented by $R \equiv d_{\text {out-of-plane}}/d_{\text {in-plane}}$ and the in-plane distortion of the MnO$_4$ plaquette represented by $d_L/d_{\text {in-plane}}$. (b) Schematic picture of the lattice distortion in the MnO$_2$ plane.}
\label{fig2}
\end{center}
\end{figure}

In order to understand the underlying mechanism of the cross-type orbital ordering, we have studied orbital ordering in La$_{0.5}$Sr$_{1.5}$MnO$_4$ in the presence of lattice distortion using unrestricted Hartree-Fock calculation on the MnO$_2$ square lattice. 
We have employed the multiband $p$-$d$ model, where the tenfold degeneracy of the Mn $3d$ spin-orbitals and the sixfold degeneracy of the O $2p$ spin-orbitals are taken into account \cite{Mizokawa2, Mizokawa3}.

The model Hamiltonian is given by
\begin{eqnarray}
H = H_p + H_d + H_{pd},
\end{eqnarray}
\begin{eqnarray}
H_p = {\displaystyle \sum_{k,l,\sigma}}{\mathstrut{
\epsilon^p_{k} p^\dagger_{kl\sigma}p_{kl\sigma}
+}} {\displaystyle \sum_{k,l>l',\sigma} 
V^{pp}_{kll'} p^\dagger_{kl\sigma}p_{kl'\sigma}}
+ \mathrm{H.c.},
\end{eqnarray}
\begin{eqnarray}
H_d & = & \epsilon_d^0 {\displaystyle \sum_{i,\alpha,m,\sigma}}
d^\dagger_{i\alpha m\sigma}d_{i\alpha m\sigma}
\nonumber \\
& + & u {\displaystyle \sum_{i,\alpha,m}}
d^\dagger_{i\alpha m\uparrow}d_{i\alpha m\uparrow}
d^\dagger_{i\alpha m\downarrow}d_{i\alpha m\downarrow}
\nonumber \\
& + & u' {\displaystyle \sum_{i,\alpha,m \neq m'}}
d^\dagger_{i\alpha m\uparrow}d_{i\alpha m\uparrow}
d^\dagger_{i\alpha m'\downarrow}d_{i\alpha m'\downarrow}
\nonumber \\
& + & (u'-j) {\displaystyle \sum_{i,\alpha,m > m',\sigma}}
d^\dagger_{i\alpha m\sigma}d_{i\alpha m\sigma}
d^\dagger_{i\alpha m'\sigma}d_{i\alpha m'\sigma}
\nonumber \\
& + & j {\displaystyle \sum_{i,\alpha,m \neq m'}}
d^\dagger_{i\alpha m\uparrow}
d_{i\alpha m'\uparrow}
d^\dagger_{i\alpha m\downarrow}
d_{i\alpha m'\downarrow}
\nonumber \\
& + & j {\displaystyle \sum_{i,\alpha,m \neq m'}}
d^\dagger_{i\alpha m\uparrow}
d_{i\alpha m'\uparrow}
d^\dagger_{i\alpha m'\downarrow}
d_{i\alpha m\downarrow},
\end{eqnarray}
\begin{eqnarray}
H_{pd} = {\displaystyle \sum_{k,l,\alpha,m,\sigma}} V^{pd}_{kl\alpha m}
d^\dagger_{k\alpha m\sigma}p_{kl\sigma} + \mathrm{H.c.}
\end{eqnarray}
Here, $d^\dagger_{i\alpha m\sigma}$ is a creation operator for the $m$-th 3$d$ electron with spin $\sigma$ at site $\alpha$ of the $i$th unit cell and $d^\dagger_{k\alpha m\sigma}$ and $p^\dagger_{kl\sigma}$ are creation operators for $m$-th 3$d$ and $l$-th 2$p$ Bloch electrons with spin $\sigma$ with wave vector $k$, respectively. $V^{pp}_{kll'}$ and $V^{pd}_{klm}$ are O 2$p$-O 2$p$ and O 2$p$-Mn 3$d$ transfer integrals. The transfer integrals are given in terms of Slater-Koster parameters ($pp\sigma$), ($pp\pi$), ($pd\sigma$) and ($pd\pi$). In the present calculation, they are fixed at ($pp\sigma$) $= 0.60$ eV, ($pp\pi$) $= -0.15$ eV, ($pd\sigma$) $= -1.8$ eV and ($pd\pi$) $= 0.81$ eV for the averaged Mn-O distance of the Mn$^{3+}$ site \cite{Mizokawa1, saitoh1, saitoh2}. Also, the atomic distance dependence of ($pd\sigma$) and ($pd\pi$) or ($pp\sigma$) and ($pp\pi$) are assumed to obey Harrison's rule $d^{-3.5}$ or $d^{-2}$ (ref. \cite{Harrison}). The $p$-$d$ overlap integrals are fixed at $S_\sigma$ $= 0.108$ eV and $S_\pi$ $= -0.0486$ eV. The intra-atomic Coulomb interaction between the Mn $3d$ electrons is expressed using Kanamori parameters, $u, u'$ and $j$ \cite{Kanamori}, which are fixed at $u=u'+2j$, $u=7.19$ eV, $u'=5.67$ eV and $j= 0.76$ eV. The charge-transfer energy $\Delta$ is defined by $\epsilon_d^0 - \epsilon_p + 4 U$ for Mn$^{3+}$, where $\epsilon_d^0$ and $\epsilon_p$ are the energies of the bare $3d$ and $2p$ orbitals and $U ( =u-20/9j)$ is the multiplet-averaged $d$-$d$ Coulomb interaction and is fixed at 4 eV. The unit cell for the CE-type AFM state thus contains eight Mn sites. The total electron number in the unit cell is 220 corresponding to the 8 Mn$^{3.5+}$ and 32 O$^{2-}$ ions. 

We have applied the Hartree-Fock approximation to the model Hamiltonian to simplify the Coulomb interaction terms to one-electron operators. For example, the second term in eq. (3) is expressed as 
\begin{eqnarray}
& u & {\displaystyle \sum_{i,\alpha,m}
d^\dagger_{i\alpha m\uparrow}d_{i\alpha m\uparrow}
d^\dagger_{i\alpha m\downarrow}d_{i\alpha m\downarrow}} 
\nonumber \\
& = & u {\displaystyle \sum_{i,\alpha,m}} \langle d^\dagger_{i\alpha m\uparrow} d_{i\alpha m\uparrow} \rangle
d^\dagger_{i\alpha m\downarrow}d_{i\alpha m\downarrow} 
\nonumber \\
& + & u {\displaystyle \sum_{i,\alpha,m}}
d^\dagger_{i\alpha m\uparrow}d_{i\alpha m\uparrow}
\langle d^\dagger_{i\alpha m\downarrow} d_{i\alpha m\downarrow} \rangle
\nonumber \\
& - & u {\displaystyle \sum_{i,\alpha,m}}
\langle d^\dagger_{i\alpha m\uparrow} d_{i\alpha m\uparrow} \rangle \langle d^\dagger_{i\alpha m\downarrow} d_{i\alpha m\downarrow} \rangle.
\end{eqnarray}
The wave function is given by a single Slater determinant. We have iterated the self-consistency cycle until successive differences of the order parameters $\langle d^\dagger_{i\alpha m \sigma}d_{i\alpha m \sigma}\rangle$ and $\langle d^\dagger_{i\alpha m \sigma}d_{i\alpha m' \sigma'}\rangle$ converged to less than 1$\times$10$^{-4}$.

We consider two parameters for lattice distortion in La$_{0.5}$Sr$_{1.5}$MnO$_4$, that is, the out-of-plane elongation ratio of the MnO$_6$ octahedra and the in-plane distortion of the MnO$_4$ plaquette around the Mn$^{3+}$ site as shown in Fig. 2(a). The former is defined by $R \equiv d_{\text {out-of-plane}}/d_{\text {in-plane}}$, where $d_{\text {out-of-plane}}$ is the distance between Mn and out-of-plane oxygen and $d_{\text {in-plane}}$ is that between Mn and in-plane oxygen. The latter is defined by $d_L/d_{\text {in-plane}}$, where $d_L$(Mn$^{3+}$) and $d_S$(Mn$^{3+}$) are the longer and shorter Mn-O distances respectively and satisfy the relationship 2$d_{\text {in-plane}}$ =  $d_L$(Mn$^{3+}$) + $d_S$(Mn$^{3+}$). The distortion in the MnO$_2$ plane is shown in Fig. 2(b).

\begin{figure}
\begin{center}
\includegraphics[width=9cm]{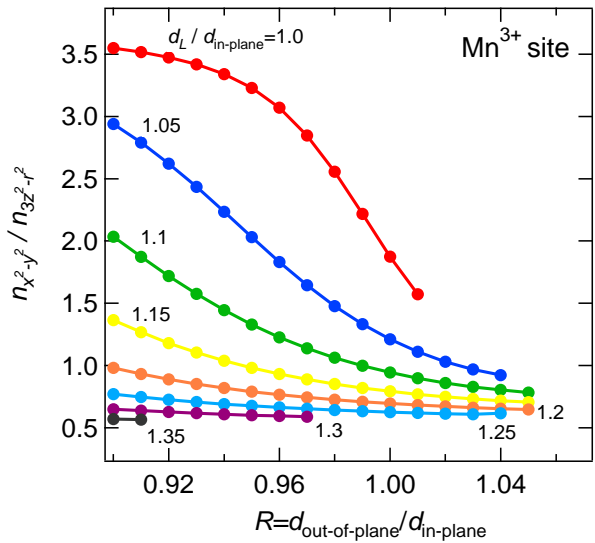}
\caption{(Color online) Orbital distribution of $e_g$ electrons at the ``Mn$^{3+}$" site.}
\label{Mn3}
\end{center}
\end{figure}

\begin{figure}
\begin{center}
\includegraphics[width=9cm]{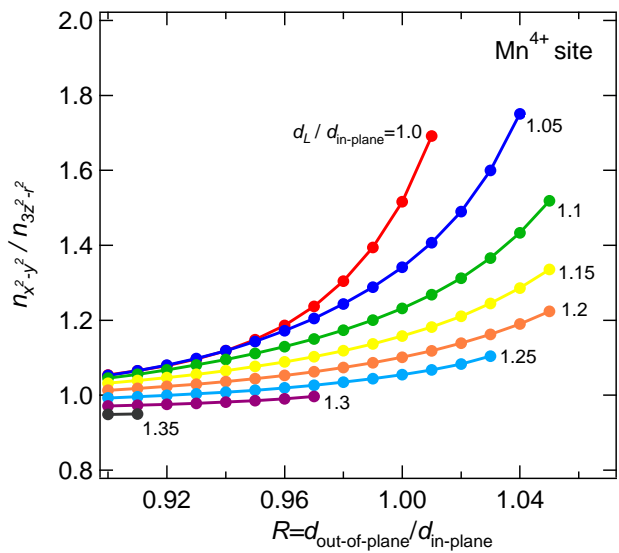}
\caption{(Color online) Orbital distribution of $e_g$ electrons at the ``Mn$^{4+}$" site.}
\label{Mn4}
\end{center}
\end{figure}

In Fig. 3, we have plotted the distribution of electrons on the different $e_g$ orbitals at the ``Mn$^{3+}$" site as a function of the lattice distortion, $R \equiv d_{\text {out-of-plane}}/d_{\text {in-plane}}$ and $d_L/d_{\text {in-plane}}$. We have selected ${x^2-y^2}$ and ${3z^2-r^2}$ as the basis orbitals for the $e_g$ states. Although the lattice distortion causes a slight tilting of the MnO$_6$ octahedra, in the present model we have neglected the tilting and included only the bond length changes for simplicity. In the limit of weak $p$-$d$ hybridization, the electron numbers of the $e_g$ orbitals $n_{x^2-y^2} = 0.75$ and $n_{3z^2-r^2} = 0.25$, that is, the relative population $n_{x^2-y^2}/n_{3z^2-r^2} = 3$ for the rod-type $(3x^2-r^2/3y^2-r^2)$, while $n_{x^2-y^2} = 0.25$ and $n_{3z^2-r^2} = 0.75$, that is, $n_{x^2-y^2}/n_{3z^2-r^2} \simeq 0.33$ for the cross-type $(z^2-x^2/y^2-z^2)$ \cite{calculation}. Therefore, one can use $n_{x^2-y^2}/n_{3z^2-r^2}$ as a measure of the type of orbital ordering. When the in-plane distortion is absent ($d_L/d_{\text {in-plane}} = 1$) and the ratio of the out-of-plane to in-plane Mn-O distances is small ($R \simeq 0.9$), the ${x^2-y^2}$ orbital is dominantly populated. As long as the in-plane distortion at the ``Mn$^{3+}$" site is weak ($d_L/d_{\text {in-plane}} \simeq 1$) and $R \alt 1.0$, the orbital distribution of $e_g$ electrons remains close to the rod-type orbital ordering. However, when the in-plane distortion becomes strong ($d_L/d_{\text {in-plane}}>1$), the distribution of $e_g$ electrons approaches that of the cross-type orbital ordering. With the in-plane distortion and the elongation along the $c$-axis, La$_{0.5}$Sr$_{1.5}$MnO$_4$ shows stronger tendency toward an orbital polarization of strong $z$ character. The dominant occupation of the ${3z^2-r^2}$ orbital causes the abrupt drop as shown in Fig. 5. In the previous HF calculation, which includes only the in-plane distortion, it has been reported that the ``Mn$^{3+}$" sites in the CE-type AFM states are accompanied by the rod-type orbital ordering \cite{Mizokawa1}. In the present case, by considering the out-of-plane elongation in addition to the in-plane distortion, we have shown that both types of distortions are important to realize the cross-type orbital ordering. We have also carried out calculations by changing the Coulomb interaction \cite{calculation2}. As a result, the relative population of the $e_g$ orbital for ``Mn$^{3+}$" changes slightly. However, the orbital ordering has also been found to be mainly determined by the effect of lattice distortion rather than that of the on-site Coulomb interaction.
\begin{figure}
\begin{center}
\includegraphics[width=9cm]{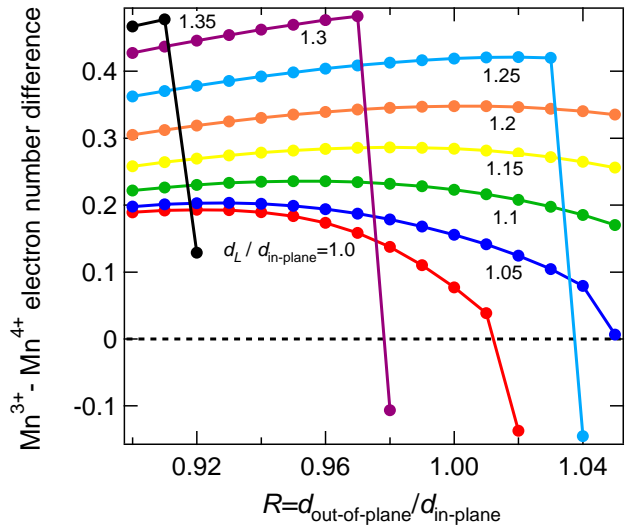}
\caption{(Color online) Difference in the number of electrons between the ``Mn$^{3+}$" and ``Mn$^{4+}$" sites.}
\label{number}
\end{center}
\end{figure}

As for the three-dimensional Mn perovskites, the resonant x-ray scattering study of Nd$_{0.5}$Sr$_{0.5}$MnO$_3$ has strongly suggested an orbital ordering of the rod-type $(3x^2-r^2/3y^2-r^2)$ \cite{Nakamura}. Because the Mn-O distance along the $c$-axis in the Nd$_{0.5}$Sr$_{0.5}$MnO$_3$ is much shorter than that within $ab$ plane, the $x^2-y^2$ orbital becomes occupied by more electrons, and the rod-type orbital ordering is favored. This is contrasted with the present case of La$_{0.5}$Sr$_{1.5}$MnO$_4$, where the in-plane and out-of-plane Mn-O distances are similarly strong, and the cross-type orbital ordering is realized.

We have also calculated the orbital distribution of $e_g$ electrons at the ``Mn$^{4+}$" site of La$_{0.5}$Sr$_{1.5}$MnO$_4$. In Fig. 4, we have plotted the orbital distribution of $e_g$ electrons at the ``Mn$^{4+}$" site as a function of lattice distortion. Here, it should be noted that there is finite number of $e_g$ electrons at the ``Mn$^{4+}$" site in spite of the formal $t_{2g}^3$ configuration because of the $p$-$d$ hybridization. For $R<1$, the orbital distribution at the ``Mn$^{4+}$" site is isotropic ($n_{x^2-y^2} \simeq n_{3z^2-r^2}$) as expected for the $t_{2g}^3$ configuration of the ``Mn$^{4+}$" ion. However, for $R>1$, the $x^2-y^2$ orbital becomes partially occupied through the anisotropic $p$-$d$ hybridization of the distortion MnO$_6$ octahedra, and the orbital distribution at the ``Mn$^{4+}$" site becomes anisotropic. In fact, because the MnO$_6$ octahedra for the ``Mn$^{4+}$" site in La$_{0.5}$Sr$_{1.5}$MnO$_4$ is elongated along the $c$-axis, the Mn-O distances within the $ab$ plane become short in comparison with that along the $c$-axis. As a result, the strong in-plane covalent bonding leads to a net charge transfer to the $x^2-y^2$ orbital. As $R$ further increases, $n_{x^2-y^2}$ strongly increases, as indicated by the sudden increase of $n_{x^2-y^2}/n_{3z^2-r^2}$ in Fig. 4. Because of the $p$-$d$ covalence, the net charge difference between ``Mn$^{3+}$" and ``Mn$^{4+}$" is significantly reduced from 1 as shown in Fig. 5. For large $R$, the difference in the number of electrons between ``Mn$^{3+}$" and ``Mn$^{4+}$" sites is reversed as shown by the sudden drop in Fig. 5. This means that the CE-type AFM state in La$_{0.5}$Sr$_{1.5}$MnO$_4$ becomes unstable for strongly elongating of the MnO$_6$ octadedra along the $c$-axis.

In conclusion, we have studied the orbital ordering in La$_{0.5}$Sr$_{1.5}$MnO$_4$ by means of unrestricted Hartree-Fock calculations. It is shown that cross-type $(z^2-x^2/y^2-z^2)$ orbital ordering originates from the in-plane distortion of the MnO$_6$ octahedra as well as the relatively long out-of-plane Mn-O distance. Further, we have found that the orbital distribution at the ``Mn$^{4+}$" site is isotropic for $R<1$, but, becomes anisotropic for $R>1$. Other physical implications of the different kinds of distortions and the orbital ordering in manganites are an open question for future studies.

Informative discussion with D. J. Huang and A. Tanaka is gratefully acknowledged. This work was supported by a Grant-in-Aid for Scientific Research in Priority Area ``Invention of Anomalous Quantum Materials" from the Ministry of Education, Culture, Sports, Science and Technology, Japan.


\begin{thebibliography}{10}
\bibitem{calculation}
In the present calculation, which takes the $p$-$d$ hybridization into consideration, it is considered that the orbital distribution of rod-type and cross-type is somewhat deviated from the ionic values. For example, $n_{x^2-y^2} \simeq 0.435$ and $n_{3z^2-r^2} \simeq 0.761$ for $R = 0.9$ and $d_L/d_{\text {in-plane}} = 1.35$. Also, non-diagonal populations $n_{x^2-y^2/3z^2-r^2}$ between the ${x^2-y^2}$ and ${3z^2-r^2}$ orbitals are as follows: $n_{x^2-y^2/3z^2-r^2} \simeq 0.027$ for $R = 0.9$ and $d_L/d_{\text {in-plane}} = 1.0$, $n_{x^2-y^2/3z^2-r^2} \simeq 0.28$ for $R = 0.9$ and $d_L/d_{\text {in-plane}} = 1.1$ and $n_{x^2-y^2/3z^2-r^2} \simeq 0.36$ for $R = 0.9$ and $d_L/d_{\text {in-plane}} = 1.35$.
\end{thebibliography}
\end{document}